# Towards an understanding of how humans perceive stiffness during bimanual exploration*

Mohit Singhala[1], Jacob Carducci[1], and Jeremy D. Brown[1]

*Abstract*— In this paper, an experimental testbed and associated psychophysical paradigm are presented for understanding how people discriminate torsional stiffness using wrist rotation about their forearm. Featured in the testbed are two 1-DoF rotary kinesthetic haptic devices. An adaptive staircase was used to evaluate JNDs for a stiffness discrimination task where participants explored virtual torsion springs by rotating their forearms. The JNDs were evaluated across seven different conditions, under four different exploration modes: bimanual, unimanual, bimanual feedback for unimanual displacement, and unimanual feedback for bimanual displacement. The discrimination results will inform future investigation into understanding how stiffness percepts vary.

## I. INTRODUCTION

Humans use a variety of exploration strategies to determine the physical and mechanical properties of objects like their shape, size, stiffness, and weight [1]–[4]. Depending on the task requirements, we may employ unimanual or bimanual strategies to evaluate the same parameters. For example, we can estimate an object's stiffness by compressing it with one hand against a hard surface like a table, or by compressing the same object between both hands. With bimanual exploration, additional consideration must be given as to whether one or both hands actively explore the object.

Even in the simple stiffness identification task described above, there are at least five different modes of exploration, resulting in five different inputs to the sensory system for the same parameters needed for percept development. From prior investigations of unimanual curvature and length discrimination, it has been demonstrated that perception may depend on the hand performing the exploration [5], [6]. Several studies have also shown non-dominant hands to be more sensitive to proprioceptive and cutaneous haptic cues [7], [8]. Similarly, bimanual and unimanual exploration modes have also been shown to result in significant perceptual differences [9], [10]. Therefore, depending on the sensory inputs, we expect the perception to change.

At present, there is a lack of consensus in the literature on the differences between unimanual and bimanual perception. There is evidence to suggest that bimanual percepts are superior to unimanual percepts due to an optimal integration of information from both limbs [11]. Likewise, there is competing evidence to suggest that bimanual percepts are formed by selecting information from one limb while neglecting information from the other [5]. Additionally, for tasks like stiffness perception, understanding how sensory information is processed will entail an investigation that goes beyond treating the two hands as individual subsystems, but rather treating each haptic sensory input as a separate variable.

It is with these considerations in mind that we introduce an experimental testbed and an associated psychophysical paradigm that aim to understand the manner in which stiffness percepts vary 1) between unimanual and bimanual exploration, and 2) when each of the four constituent sensory inputs – displacements or torque feedback from each hand – are selectively removed. In the following sections, we describe the experimental testbed and psychophysical paradigm as well as results from a pilot experiment in which they were employed.

*This material is based upon work supported by the National Science Foundation under NSF Grant# 1657245

[1]Mohit Singhala, Jacob Carducci, and Jeremy D. Brown are with the Department of Mechanical Engineering, Johns Hopkins University, Baltimore, MD 21218, USA `mohit.singhala@jhu.edu`

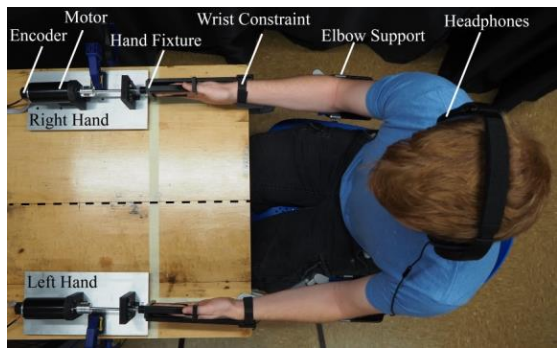

Fig. 1. Experimental setup for bimanual stiffness exploration. The left hand setup is identical to the right hand setup.

## II. METHODS

We recruited n = 3 individuals (3 male, age = 23 ± 2 years) to distinguish virtual torsional springs with different spring constants. The participants were compensated at a rate of $10/hour. All participants were consented according to a protocol approved by the Johns Hopkins School of Medicine Institutional Review Board (Study# IRB00148746).

### A. Apparatus

Our experimental apparatus consists of two identical custom 1-DoF rotary kinesthetic haptic devices. Each device features a Maxon RE50 motor (200 W) equipped with a 3-channel 500 CPT encoder. The motors are driven by a Quansar AMPAQ-L4 linear current amplifier and can achieve a maximum peak torque of 467 mNm. Control is provided through a Quanser QPIDe data acquisition card with a

MATLAB/Simulink and QUARC interface run at a frequency of 1 kHz.

*B. Procedure*

In this study, participants explored virtual springs by rotating their wrists about their forearms in an two-interval two-alternative forced-choice (2I-2AFC) stiffness discrimination task. The Just Noticeable Difference (JND) of torsional stiffness was determined using the adaptive staircase algorithm described below.

Participants interacted with the setup through custom 3D-printed hand fixtures enabling coaxial rotation of their wrist(s) with the motor shaft. Elbow support and a height-adjustable chair were used to ensure the participant's elbows were kept as close to 90 degrees as possible with minimal radial and ulnar deviation. The orientation of the participant's wrists at the beginning of each trial is shown in Fig. 1. This position is henceforth referred to as the neutral position. Participants pronated their wrists within their range of motion and returned to neutral position without repeated exploration of the same stiffness.

For each trial, a pair of virtual springs – test and reference – were presented sequentially in a random order. Identical audio cues were provided to the participant and the experimenter using separate pairs of noise-cancelling headphones, and notified the participants if they were exploring the first or second spring. Participants were instructed to respond with their choice of which spring felt stiffer for each trial. A single reference spring was used with a spring constant of 1.5 mNm/deg for the entire experiment. The spring constant for the test spring was determined by the staircase algorithm.

Once a participant's response was recorded, the stiffness of the next test spring was scaled. This stiffness scaling was adjusted based on a 1-up-3-down transformed weighted staircase algorithm [12]. Each staircase started with the scaling factor of the test spring at 150% of the reference spring, and with upward steps of 10% and downward steps of 7.32%. After two reversals, the upward steps changed to 5% and the downward steps changed to 3.66%. Consequentially, the proportion-correct target of the algorithm was 83.15%. The staircase terminated after ten reversals, and the discrimination JND was obtained by averaging the stiffness values of the last eight reversals. An example of this adaptive staircase algorithm for a given condition is shown in Fig. 2.

Participants were instructed to direct their visual focus away from the setup while exploring the springs. During exploration, the reaction torque rendered by the motors was

$$\tau_L = \tau_R = s \cdot \kappa \cdot (\theta_L + \theta_R) \quad (1)$$

where $\tau_L$ and $\tau_R$ are the torques generated on the left and right wrists, $s$ is the scaling factor of the spring, $\kappa$ is the stiffness of the reference spring, and $\theta_L$ and $\theta_R$ are the angular displacements of the left and right wrists. For the reference spring $s = 1$, whereas for the test spring $s \geq 1$.

*C. Conditions*

The experiment consisted of seven conditions representing the four major modes of exploration described in Table I.

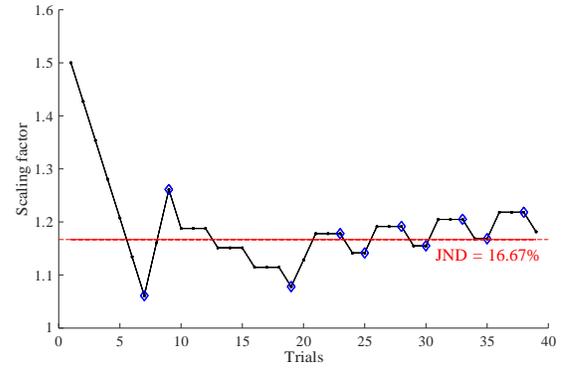

Fig. 2. Adaptive staircase from representative participant for determining estimate of stiffness discrimination threshold. Trial markers with blue diamonds indicate reversals. Labeled JND is from this specific staircase.

TABLE I
PRESENTATION TYPES AND SIDE VARIANTS

| Condition | Rendering | Diagram |
|---|---|---|
| 1 | $\tau_L = \tau_R = s \cdot \kappa \cdot (\theta_L + \theta_R)$ | |
| 2L | $\tau_L = s \cdot \kappa \cdot \theta_L$, $\tau_R = 0$ | |
| 2R | $\tau_R = s \cdot \kappa \cdot \theta_R$, $\tau_L = 0$ | |
| 3L | $\tau_L = \tau_R = s \cdot \kappa \cdot \theta_R$ | |
| 3R | $\tau_L = \tau_R = s \cdot \kappa \cdot \theta_L$ | |
| 4L | $\tau_R = s \cdot \kappa \cdot (\theta_L + \theta_R)$, $\tau_L = 0$ | |
| 4R | $\tau_L = s \cdot \kappa \cdot (\theta_L + \theta_R)$, $\tau_R = 0$ | |

- In Condition 1, participants performed bimanual exploration with both wrists. The rendering of Eqn. 1 determined the reaction torques imparted on the exploring wrists.
- In Condition 2L, participants performed unimaual exploration with their left wrist. The other end of the virtual spring was rendered as a grounded end.
- Condition 2R is the right-wrist version of 2L.
- In Condition 3L, participants performed bimanual exploration keeping their left wrist in the neutral position while exploring the spring with their right wrist. Any displacement from the left wrist was not included in stiffness rendering.
- Condition 3R is the right-wrist version of 3L.
- In Condition 4L, participants performed bimanual ex-

ploration with both wrists, but the virtual spring's reaction torque was not displayed to the left wrist.
- Condition 4R is the right-wrist version of 4L.

Each participant completed one staircase for each of the seven conditions with the presentation of the four modes of exploration randomized. For Conditions 2-4, the presentation of the left and right variants were randomized as well. A five-minute break was provided after each condition and participants took an average of 105 minutes to finish the entire experiment. After the experiment, participants answered questions from the Edinburgh Handiness Survey.

TABLE II
JUST NOTICEABLE DIFFERENCES (JND) (%) OF STIFFNESS DISCRIMINATION

|  | Participants | | |
| --- | --- | --- | --- |
| Conditions | S1 | S2 | S3 |
| 1 | 15.69 | 12.63 | 17.43 |
| 2L | 16.93 | 12.58 | 8.23 |
| 2R | 12.70 | 15.95 | 14.30 |
| 3L | 14.46 | 8.37 | 16.98 |
| 3R | 19.46 | 14.57 | 8.08 |
| 4L | 21.24 | 9.26 | 29.92 |
| 4R | 20.13 | 16.67 | 15.53 |

## III. RESULTS & DISCUSSION

All participants in the experiment were right-hand dominant.

The preliminary experiment helped demonstrate the feasibility of performing an entire protocol in a single session with an average time of 105 minutes, thereby avoiding possible confounds from a multi-day study. The JND values for all seven conditions from the three participants are shown in Table II. All JND values were found to be within expected ranges based on values reported in the literature [3], [13]. Given the low sample size of the preliminary experiment, no statistical analyses were performed. However, we noticed an interesting observation where all three participants discriminated better when either wrist was held stationary compared to when torque feedback was not provided to that same wrist (Conditions 3 versus 4). While we realize that the sample is too small to form any conclusions, these observations and the potential of stationary positioning being more important for constructing stiffness percepts warrant a full-scale study, with a large enough sample size to allow for meaningful statistical analyses.

## IV. CONCLUSION

To address the gap in research on the formation of stiffness perception from the wrists, we introduced an experimental setup and protocol to examine perception as a function of hand dominance and information presentation. Our findings highlight the potential of this setup to investigate the impact of both unimanual and bimanual modes of exploration, as well as the utility of constituent sensory inputs, on stiffness discrimination within a single testing session. These preliminary observations motivate further investigation with sufficient statistical power. An improved understanding of stiffness perception can motivate the design of haptic interfaces that consider fully the manner in which haptic exploration affects our perception.

## ACKNOWLEDGMENT

This material is based upon work supported by National Science Foundation under NSF Grant # 1657245.

## REFERENCES

[1] R. L. Klatzky, S. J. Lederman, and V. A. Metzger, "Identifying objects by touch: An "expert system"," *Perception & Psychophysics*, 1985.
[2] H. Z. Tan, B. D. Adelstein, R. Traylor, M. Kocsis, and E. D. Hirleman, "Discrimination of real and virtual high-definition textured surfaces," in *14th Symposium on Haptics Interfaces for Virtual Environment and Teleoperator Systems 2006 - Proceedings*, 2006.
[3] L. A. Jones and I. W. Hunter, "A perceptual analysis of stiffness," *Experimental Brain Research*, vol. 79, no. 1, pp. 150–156, 1990.
[4] J. D. Brown, M. K. Shelley, D. Gardner, E. A. Gansallo, and R. B. Gillespie, "Non-Colocated Kinesthetic Display Limits Compliance Discrimination in the Absence of Terminal Force Cues," *IEEE Transactions on Haptics*, vol. 9, no. 3, pp. 387–396, 2016.
[5] V. Squeri, A. Sciutti, M. Gori, L. Masia, G. Sandini, and J. Konczak, "Two hands, one perception: How bimanual haptic information is combined by the brain," *Journal of Neurophysiology*, vol. 107, no. 2, pp. 544–550, 2012.
[6] V. Panday, W. M. Bergmann Tiest, and A. M. Kappers, "Bimanual and unimanual length perception," *Experimental Brain Research*, vol. 232, no. 9, pp. 2827–2833, 2014.
[7] D. J. Goble and S. H. Brown, "Upper limb asymmetries in the matching of proprioceptive versus visual targets," *Journal of Neurophysiology*, 2008.
[8] D. J. Goble, B. C. Noble, and S. H. Brown, "Proprioceptive target matching asymmetries in left-handed individuals," *Experimental Brain Research*, 2009.
[9] A. M. Kappers and J. J. Koenderink, "Haptic unilateral and bilateral discrimination of curved surfaces," *Perception*, 1996.
[10] A. F. Sanders and A. M. Kappers, "Bimanual curvature discrimination of hand-sized surfaces placed at different positions," *Perception and Psychophysics*, vol. 68, no. 7, pp. 1094–1106, 2006.
[11] M. A. Plaisier and M. O. Ernst, "Two hands perceive better than one," in *Lecture Notes in Computer Science (including subseries Lecture Notes in Artificial Intelligence and Lecture Notes in Bioinformatics)*, 2012.
[12] M. A. García-Pérez, "Forced-choice staircases with fixed step sizes: Asymptotic and small-sample properties," *Vision Research*, vol. 38, no. 12, pp. 1861–1881, 1998.
[13] A. Paljic, J. . Burkhardtt, and S. Coquillart, "Evaluation of pseudo-haptic feedback for simulating torque: a comparison between isometric and elastic input devices," in *12th International Symposium on Haptic Interfaces for Virtual Environment and Teleoperator Systems, 2004. HAPTICS '04. Proceedings.*, March 2004, pp. 216–223.